\documentclass[iop,apjl]{emulateapj}

\newcommand{\simless}{\mathbin{\lower 3pt\hbox
      {$\rlap{\raise 5pt\hbox{$\char'074$}}\mathchar"7218$}}} 
\newcommand{\simgreat}{\mathbin{\lower 3pt\hbox
     {$\rlap{\raise 5pt\hbox{$\char'076$}}\mathchar"7218$}}} 

\slugcomment{\today}

\shorttitle{VLA observations of 2M0444}
\shortauthors{Ricci et al.~(2017)}

\begin{document}


\title{VLA Observations of the disk around the young brown dwarf 2MASS J044427+2512}


\author{L. Ricci}

\affil{Department of Physics and Astronomy, Rice University, 6100 Main Street, 77005 Houston, TX, USA}

\and

\author{H. Rome}
\affil{The Kinkaid School, 201 Kinkaid School Dr, 77024 Houston, TX, USA}

\and

\author{P. Pinilla}
\affil{Department of Astronomy Steward Observatory, The University of
Arizona, 933 North Cherry Avenue, 85721 Tucson, AZ, USA}

\and

\author{S. Facchini}
\affil{Max-Planck-Institut fur Extraterrestrische Physik, Giessenbachstrasse 1, 85748 Garching, Germany}

\and

\author{T. Birnstiel}
\affil{University Observatory, Faculty of Physics, Ludwig-Maximilians-Universit\"at M\"unchen, Scheinerstr. 1, 81679 Munich, Germany}

\and

\author{L. Testi}
\affil{European Southern Observatory (ESO) Headquarters, Karl-Schwarzschild-Str. 2, 85748 Garching, Germany}

\email{luca.ricci@rice.edu}


\begin{abstract}

We present multi-wavelength radio observations obtained with the VLA of the protoplanetary disk surrounding the young brown dwarf 2MASS J04442713+2512164 (2M0444)  in the Taurus star forming region. 2M0444 is the brightest known brown dwarf disk at millimeter wavelengths, making this an ideal target to probe radio emission from a young brown dwarf. Thermal emission from dust in the disk is detected at 6.8 and 9.1 mm, whereas the 1.36 cm measured flux is dominated by ionized gas emission.   
We combine these data with previous observations at shorter sub-mm and mm wavelengths to
test the predictions of dust evolution models in gas-rich disks after adapting their parameters to the case of 2M0444. These models show that the radial drift mechanism affecting solids in a gaseous environment has to be either completely made inefficient, or significantly slowed down by very strong gas pressure bumps in order to explain the presence of mm/cm-sized grains in the outer regions of the 2M0444 disk. We also discuss the possible mechanisms for the origin of the ionized gas emission detected at 1.36 cm. The inferred radio luminosity for this emission is in line with the relation between radio and bolometric luminosity valid for for more massive and luminous young stellar objects, and extrapolated down to the very low luminosity of the 2M0444 brown dwarf.
    

\end{abstract}



\section{Introduction}
\label{sec:intro}

Young pre-Main Sequence (PMS) stars and brown dwarfs (BDs) are orbited by disks which are the cradles of planets \citep[see][for a recent review]{Andrews:2015}. According to the core accretion scenario, planets are formed via the growth of solid particles in these young circumstellar disks \citep[e.g.,][]{Mordasini:2010}. Studying the properties and growth of dust particles in these disks is therefore crucial to understand the process of planet formation. In particular, the radial drift problem for pebbles in the outer disk regions and larger rocks in the inner disk is one of the most compelling impediments to our understanding of the formation of planetesimals \citep{Weidenschilling:1977}, and significant efforts are being made on both observational and theoretical grounds \citep[e.g.,][]{Testi:2014,Johansen:2014}. 

Because of their low-density and temperature, BD disks are particularly interesting as they allow to test the models of dust evolution in borderline environments \citep{Pinilla:2013, Meru:2013}.
Several disks around young BDs, with masses between those of stars and planets, have been characterized at infrared and sub-millimeter wavelengths \citep[e.g.][]{Natta:2001,Klein:2003,Scholz:2006,Morrow:2008,Furlan:2011,Harvey:2012,Alves de Oliveira:2013,Ricci:2014,vanderPlas:2016,Testi:2016,Ricci:2017}. The study of BD disks is particularly relevant also to investigate the potential of finding exoplanets around more evolved BDs \citep{Payne:2007,Ricci:2014}. 

Here we present Karl G. Jansky Very Large Array (VLA) observations of the 2MASS J04442713+2512164 (henceforth 2M0444) system, made of a young BD \citep[M7.25-spectral type, sub-stellar luminosity $L_{\rm{BD}} \approx 0.028~L_{\odot}$, effective temperature $T_{\rm{eff}} \approx 2828$ K, and mass $M_{\rm{BD}} \approx 60~M_{\rm{Jup}}$,][]{Luhman:2004}, surrounded by a disk first detected in the infrared~\citep{Kenyon:1994,Hartmann:2005,Luhman:2006,Guieu:2007,Bouy:2008}. 2M0444 is a member of the Taurus star forming region, with an estimated age of $\approx 1 - 3$ Myr, and a distance of about 140 pc \citep{Loinard:2007,Torres:2009,Torres:2012}.

The 2M0444 disk is the brightest BD disk from the single-dish survey in Taurus by \citet{Scholz:2006} at a wavelength of 1.3 millimeter. Thanks to its relatively high flux density ($F_{\rm{1.3mm}} \approx 5$ mJy) and large disk (outer radius $> 100$ au), this was the first BD disk to be spatially resolved at mm-wavelengths~\citep[][using the Combined Array for Research in Millimeter-wave Astronomy, CARMA]{Ricci:2013}.

Subsequent observations with the Atacama Large Millimeter/submillimeter Array (ALMA) revealed a low value of the spectral index between 0.89 and 3.2 mm, $\alpha_{\rm{0.89-3.2mm}} \approx 1.8$~\citep[$F_{\nu} \propto \nu^{\alpha}$,][]{Ricci:2014}. The inferred value of the spectral index of the dust opacity $\beta$ ($\kappa_{\nu} \propto \nu^{\beta}$) at these wavelenghts indicates the presence of mm-sized grains in the outer regions of this BD disk, similarly to what is found in circumstellar disks \citep{Testi:2014}.  
 
Our new VLA observations for the continuum emission of 2M0444 at 6.8, 9.1 mm and 1.36 cm, allow us to probe dust thermal emission at wavelengths longer than $\approx$ 3 mm for the first time in a BD disk. This lets us study dust particles that are larger than can be probed with ALMA. 
Furthermore, observations at these long wavelengths can pick up emission from non-dust emission processes, and related to the activity of the young central object~\citep[e.g.,][]{Shirley:2007}.

Section~\ref{sec:obs} describes the new VLA observations and data reduction, Section~\ref{sec:results} presents the results of the observations. Section~\ref{sec:analysis} describes the analysis of the results. The main results of this work are summarized in Section~\ref{sec:conclusions}. 




\section{VLA Observations and Data Reduction}
\label{sec:obs}

We observed the 2M0444 young brown dwarf with the VLA in bands Q (effective wavelength of 6.8 mm), Ka (9.1 mm), and K (1.36 cm). The VLA correlator was configured to record dual polarization with 64 separate spectral windows, each with a total bandwidth of 128 MHz.

Observations were performed during the month of March 2013 (see Table~\ref{table:obs}), when the VLA was in the D array configuration with 27 available antennas. Baseline lengths ranged between about 35 m and 1.0 km.


The visibility datasets were calibrated with the VLA Calibration Pipeline\footnote{https://science.nrao.edu/facilities/vla/data-processing/pipeline} developed by the National Radio Astronomy Observatory (NRAO) within the CASA software package \citep{McMullin:2007}.
During the observations, the quasars J0319$+$4130, J0431$+$2037, and J0542$+$4951 were used to calibrate the frequency-dependent bandpass, gain terms and absolute flux scale, respectively. 
Observations for each band were repeated twice. 

For imaging the interferometric visibilities we used the \texttt{clean} algorithm in CASA. Since the VLA D array configuration does not allow us to spatially resolve the emission from the 2M0444 disk, we performed the imaging using a natural weighting, which maximizes sensitivity. 
The flux values extracted in the two days were generally found to be consistent within 10$\%$. We adopt this value as the uncertainty on the absolute flux calibration. The observations were then combined to produce a single final image for each band. The inferred rms and flux density values are listed in Table~\ref{table:obs}. The flux density values were extracted by integrating the disk surface brightness over circular apertures with radii of $\approx 5''$ centered on the location of 2M0444.

\begin{deluxetable}{lrcrc}[t!]  
\centering
  \tablecaption{\label{table:obs}VLA Observations of 2M0444}
  \tablehead{\colhead{{\sc Band}} & {\sc $\lambda$ [$\rm{mm}$]} & {\sc $\rm{Rms}$ [$\rm{\mu Jy/beam}]$} & {\sc $F_{\nu}$ [$\rm{\mu Jy}]$} & {\sc Observing days}}
  \startdata
  Q         &   6.8   &  6.0  & 159 & 2013 March 11, 13   \\ 
  Ka       &   9.1   &  2.8  &   71 & 2013 March 30, 31    \\
  K         &   13.6 &  2.9  &   64 & 2013 March 10, 12  \vspace{1mm} 
  \enddata
  
\end{deluxetable}

%

\section{Results}
\label{sec:results}

The 2M0444 disk was detected at all the three VLA bands. The signal-to-noise ratios are $\approx 26, 25, 22$ at $\lambda = 6.8, 9.1, 13.6$ mm, respectively (Table~\ref{table:obs}). 

\begin{figure}[t!]
\includegraphics[scale=0.5]{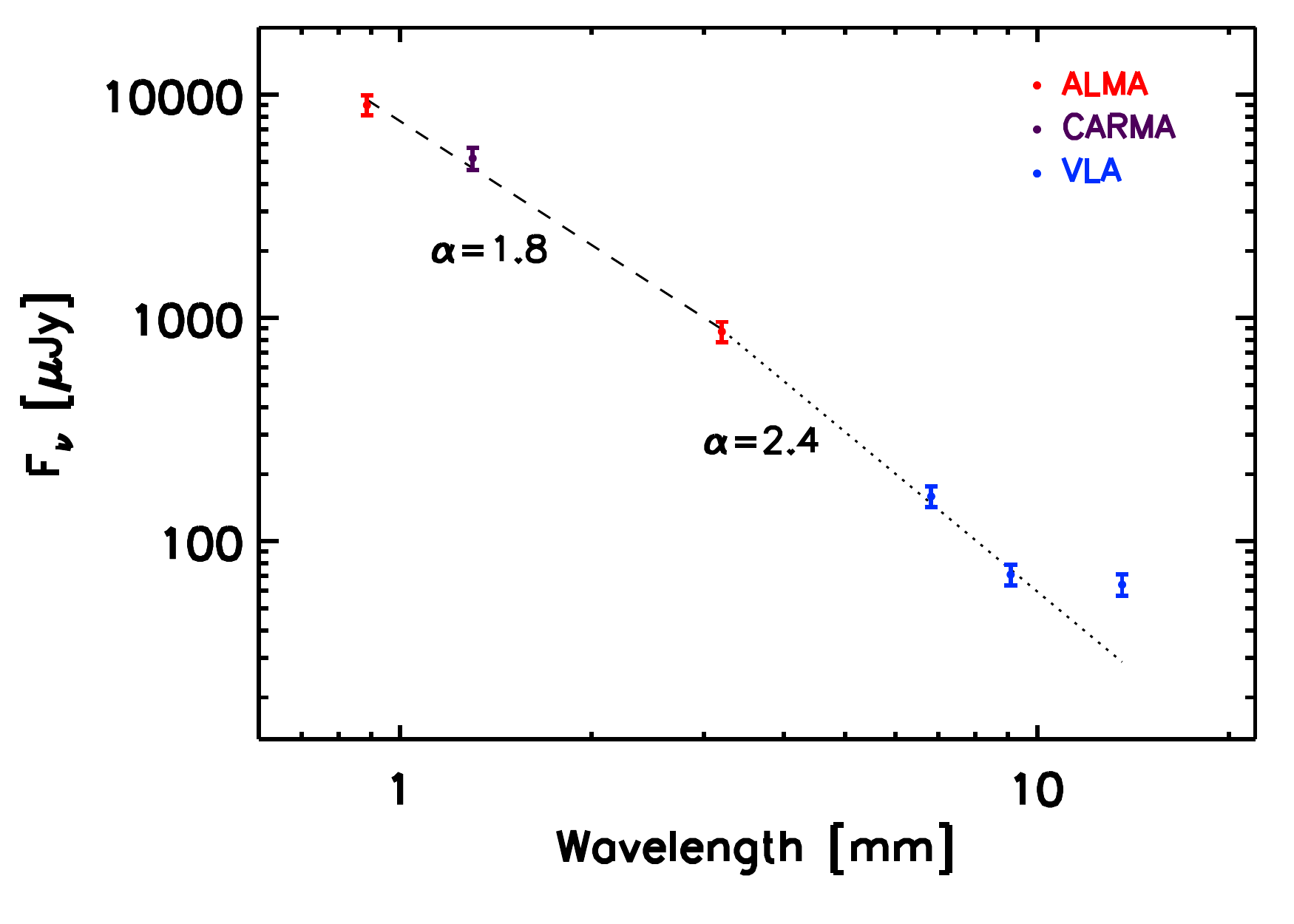}   
\vspace{-5mm}
\caption{Spectral Energy Distribution at sub-mm to cm wavelengths for 2M0444. For each datapoint, the vertical errorbar accounts for both the measured rms noise and a fluxcale absolute uncertainty of $10\%$, these two terms added in quadrature. Dashed and dotted lines show the power-law fits for wavelengths of $0.89 - 3.2$ mm and $3.2 - 9.1$ mm, respectively.}
              \label{fig:2m0444_sed}
\end{figure}

Figure~\ref{fig:2m0444_sed} displays the disk SED at sub-mm to cm wavelengths after combining the new VLA detections presented here at 6.8, 9.1 mm and 1.36 cm with previous detections with ALMA at 0.89 and 3.2 mm \citep{Ricci:2014} and CARMA at 1.3 mm \citep{Ricci:2013}. We do not include here the fluxes measured at very similar wavelengths by \citet{Scholz:2006} and \citet{Bouy:2008} because of the much lower signal-to-noise ratios than the ALMA and CARMA detections. 

This figure shows an apparent steepening of the disk SED at longer wavelengths, with evidence of \textit{excess emission} at 1.36 cm over a power-law derived fitting the flux densities measured at shorter wavelengths. 
When fitting all the flux densities measured between 0.89 and 9.1 mm with a single power-law, we infer a spectral index $\alpha_{\rm{0.89-9.1mm}} = 2.07 \pm 0.12$ ($F_{\nu} \propto \nu^{\alpha}$). However, the slope at $\lambda \simgreat 3.2$ mm is steeper than at shorter wavelengths. If fitting separately the flux densities at $0.89 - 3.2$ mm and at $3.2 - 9.1$ mm, we get $\alpha_{\rm{0.89-3.2mm}} = 1.84 \pm 0.26$, and $\alpha_{\rm{3.2-9.1mm}} = 2.37 \pm 0.32$, respectively.

The extrapolated flux density at 1.36 cm from the power-law derived between 3.2 and 9.1 mm is $\approx 29~\mu$Jy. The measured flux density of 2M0444 at this wavelength is\footnote{The reported uncertainty includes both the statistical uncertainty given by the rms noise and the $10\%$ systematic uncertainty on the absolute flux scale (Section~\ref{sec:obs}), these terms added in quadrature.} $64 \pm 7~\mu$Jy, which is $5.7\sigma$ above the power-law extrapolated value. 

\section{Analysis}
\label{sec:analysis}

\subsection{The steepening of the SED at long wavelengths}
\label{sec:steepening}

Because of the decrease of the dust opacity with wavelength \citep[e.g.,][]{Draine:2006}, the dust thermal emission from young disks around PMS stars and brown dwarfs is found to be mostly optically thin at wavelengths close to 1 mm, and longer \citep[e.g.,][]{Testi:2003,Rodmann:2006,Ricci:2012, Ricci:2013,Testi:2014}. In the optically thin regime, the disk flux density depends on frequency via $F_{\nu} \propto \kappa_{\nu} \times B_{\nu}(T_{\rm{dust}})$, where $B_{\nu}(T_{\rm{dust}})$ is the Planck function associated to the dust emission. Therefore, the spectral index of the flux density $\alpha \approx \beta + \alpha_{Pl}$, where $\beta$ and $\alpha_{\rm{Pl}}$ are the spectral indices of the dust opacity and Planck function, respectively.

\citet{Ricci:2014} constrained the structure of the 2M0444 disk by modeling the interferometric visibilities obtained with ALMA at 0.89 mm and the measured spectral index of $\approx 1.8$ between 0.89 and 3.2 mm. From this analysis, they inferred values of $\beta \approx 0.2$ and $\alpha_{\rm{Pl}} \approx 1.6$. The value of $\alpha_{\rm{Pl}} < 2$ indicates that the mm-wave emission is dominated by dust that is cold enough to depart from the Rayleigh-Jeans regime of the emission. In particular, the \citet{Ricci:2014} models predict a temperature radial profile decreasing with the distance from the 2M0444 brown dwarf, with values of $\approx 10 - 13$ K between 40 au and the outer radius of the disk at $\approx 140$ au.
In these models, the dust temperature in the disk outer regions is dominated by external heating due to the interstellar radiation impinging on the disk. A value of $T_{\rm{ext}} = 10$ K was assumed based on the temperature inferred for the regions of protostellar cores directly illuminated by the interstellar radiation in the Taurus region. In the appendix of the current paper we provide the results of physical radiative transfer models that support this assumption.

\begin{figure}[th!]
\includegraphics[scale=0.5]{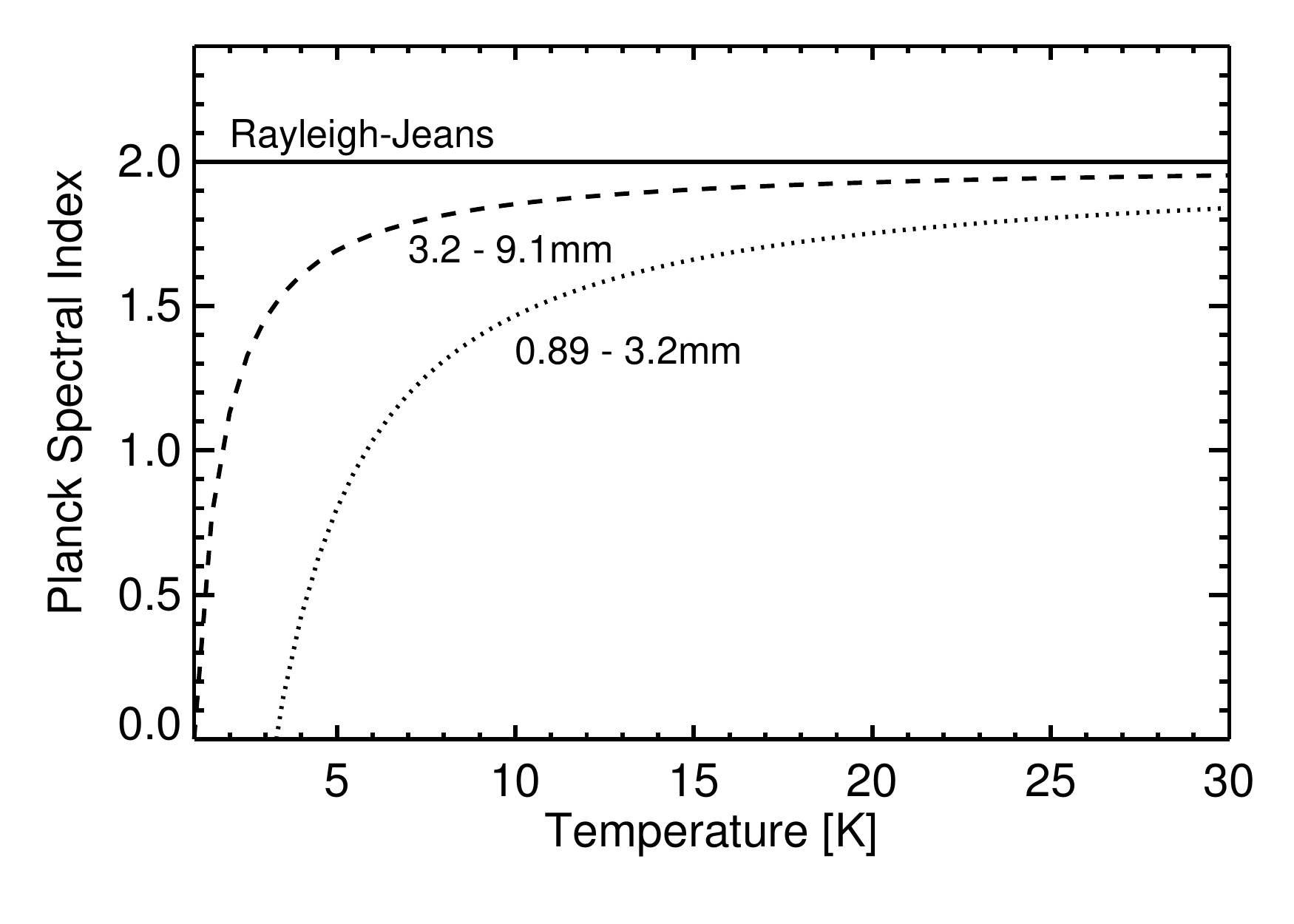}   
\vspace{-5mm}
\caption{Spectral index $\alpha_{\rm{Pl}}$ of the Planck function as a function of temperature. In this plot the x-axis represents possible temperature values of dust in the outer regions of the 2M0444 disk.  The dashed and dotted lines are for spectral indices calculated between $3.2 - 9.1$ mm and $0.89 - 3.2$ mm, respectively. The solid line indicates the value of 2 obtained in the Rayleigh-Jeans regime.}
              \label{fig:spindex_planck}
\end{figure}

As shown in Figure~\ref{fig:spindex_planck}, at a given temperature, the spectral index of the Planck function increases at longer wavelengths. For the range given above for the outer regions of the 2M0444 disk, i.e. $T = 10 - 13$ K, $\alpha_{\rm{Pl}} \approx 1.8 - 1.9$ between 3.2 and 9.1 mm. Hence, within the uncertainties on the inferred value of the SED spectral index $\alpha$ (Section~\ref{sec:results}), the steepening of the SED of the 2M0444 disk can be explained by the steepening of the Planck function at longer wavelengths as due to the low dust temperatures in the disk outer regions according to the \citet{Ricci:2014} models. 

Note that this is under the assumptions that i) the emission is optically thin, and ii) the spectral index $\beta$ of the dust opacity does not vary significantly between 0.89 and 9.1 mm. The low optical depth of the dust emission at 0.89 mm, and therefore also at longer wavelengths due to the decrease of the dust opacity with wavelength, has been confirmed by the modeling presented in \citet{Ricci:2014}. 
Regarding the variation of the dust opacity with wavelength, dust models with different assumptions on the chemical composition and grain size distribution show that $\beta$ can vary with wavelength in the spectral interval discussed here \citep[e.g., see upper panels in Figures 3-5 and Fig. 6 in][]{Draine:2006,Miyake:1993}. In the vast majority of cases, $\beta$ \textit{increases} with wavelength, in line with the steepening of the SED of the 2M0444 disk, but there are exceptions \citep{Draine:2006}.

To summarize, although the observed steepening of the SED of the 2M0444 disk can be explained solely by the steepening of the Planck function at long wavelengths for dust thermal emission at $\rm{T} \approx 10-13$ K \citep{Ricci:2014}, the predictions of dust models suggest that some contribution is likely due also to a steepening of the spectral index $\beta$ of the dust opacity at $\lambda > $ 1 mm. 

\subsection{Models of dust evolution in the 2M0444 disk}
\label{sec:dust}

The measured values of the spectral index of the disk SED at sub-mm/mm wavelengths imply\footnote{From the relation $\alpha = \beta + \alpha_{\rm{Pl}}$, in order to reproduce the measured value of $\alpha$ between 3.2 and 9.1 mm with $\beta > 1$ would require $\alpha_{\rm{Pl}} \lesssim 1.4$, given by unreasonably cold dust ($T \lesssim 3$ K, Fig.~\ref{fig:spindex_planck}, see also the appendix, Fig.~\ref{fig:r_temp}).} $\beta < 1$, which can be obtained only   by invoking dust grains as large as $\sim 1$ mm, or larger, in the outer regions of the 2M0444 disk \citep{Draine:2006,Testi:2014}. 

We compare this result with the predictions of dust evolution models that compute the dynamics of dust particles simultaneously with their growth, erosion, and fragmentation, as presented in \citet{Birnstiel:2010}. For this comparison, we adopted the (sub-)stellar properties constrained for 2M0444, i.e. a bolometric luminosity $L_{\rm{BD}} = 0.028~L_{\odot}$, effective temperature $T_{\rm{eff}} = 2838$ K \citep{Luhman:2004}, mass of $M_{\rm{BD}} = 50~M_{\rm{Jup}}$ obtained using the \citet{Baraffe:2003} evolutionary models, and similar to the value of $45~M_{\rm{Jup}}$ inferred by \citet{Bouy:2008}. For the disk, we adopted the surface density radial profile constrained by the \citet{Ricci:2014} modeling of the ALMA visibilities, i.e. a power-law function $\Sigma \propto r^{-p}$ with an exponent $p = 1.65$, truncated at an outer radius of 139 au, and with a total disk mass (gas$+$dust) of 1.3 $M_{\rm{Jup}}$ (gas-to-dust mass ratio of 100). The temperature radial profile is also taken by \citet{Ricci:2014}. We assume an $\alpha_{\rm{v}}$-disk viscosity, with a value of $10^{-3}$ \citep{Pringle:1981}.

\begin{figure*}[th!]
\centering
\includegraphics[scale=0.45]{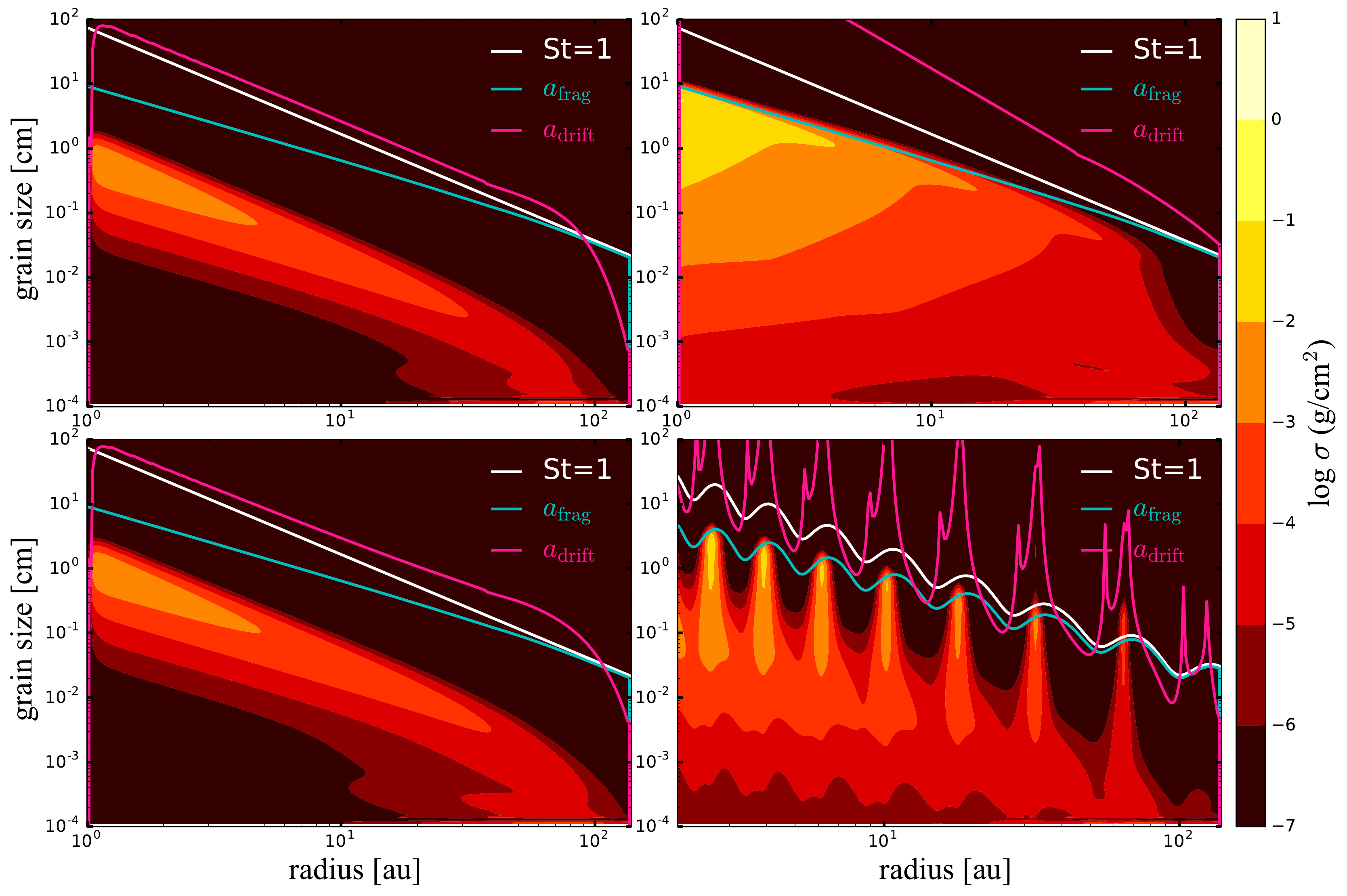}   
\vspace{-2mm}
\caption{Grain size distributions as a function of the distance from the 2M0444 for dust evolution models discussed in Section~\ref{sec:dust}. At each point, the color defines the surface density of particles with a given grain size (y-axis) and at a given distance from 2M0444 (x-axis; see colorbar on the right side). In each panel, the white solid line identifies the grain sizes with the Stokes parameter equal to 1, which are the particles most affected by radial drift; the cyan solid line shows the fragmentation barrier; the purple solid line shows the largest grain sizes that can be reached before radial drift removes those particles \citep[e.g.,][]{Birnstiel:2012}. Each panel shows a snapshot of our evolutionary models taken at $\approx 1$ Myr since the beginning of the simulations. Top left panel) Model with unimpeded radial drift. Top right) Model where radial drift is neglected. Bottom left) Model where radial drift velocities are reduced by a factor $f = 0.6$. Bottom right) Model with pressure bumps with amplitude $A = 0.6$ and wavelength equal to the local pressure scale-height.}
              \label{fig:dust_evolution}
\end{figure*}

In our numerical models we assume that the initial grain size is 1\,$\mu$m in the entire disk and with an initially constant dust-to-gas mass ratio of 0.01. In these models, particles can stick when the relative velocities are below a velocity threshold, otherwise the collision leads to erosion or complete fragmentation of particles. The relative velocities are calculated according to the radial drift, dust settling, Brownian and turbulent motions. The velocity threshold, or \textit{fragmentation velocity} $v_{\rm{frag}}$, is assumed to be 10 and 30\,m\,s$^{-1}$, in agreement with laboratory experiments and numerical simulations of collisions of icy grains \citep{Blum:2008,Gundlach:2011}.
For the dust dynamics, we consider radial drift, gas drag, and turbulence; all of these mechanisms depend on the grain size and hence the importance of modeling grain growth and dust dynamics simultaneously. We calculate the evolution up to 5\,Myr.

\citet{Pinilla:2013} demonstrated that, for disks with gas surface densities monotonically decreasing with stellocentric radius, the timescale of radial drift for $\sim$ mm-sized grains is \textit{shorter} around stars with lower mass, and it can be a severe problem to explain mm-grains in the outer regions of BD disks. For this reason, we assume in some of  these models that either the radial drift is completely neglected or reduced by a given factor ($v_{\rm{drift, reduced}}=f\times v_{\rm{drift}}$, with $f=0.4, 0.6$). In addition, we also consider models where the the gas surface density is perturbed by a sinusoidal function with certain amplitude and frequency, to mimic bumps in the pressure radial profile \citep[][]{Pinilla:2012a} which can be produced by different physical mechanisms \citep[e.g. zonal flows, dead zones, or planet-disk interaction, see][]{Johansen:2009,Uribe:2011,Pinilla:2012b}. For the sinusoidal perturbation, we consider two values for the amplitude ($A=0.4$ and 0.6) and a wavelength equal to the disk local scale height. These values are known to cause a slowing down or even halting locally the radial motion of mm-sized solids in the disk outer regions \citep[][]{Pinilla:2013}.

The results of our calculations are shown in Fig.~\ref{fig:dust_evolution}, which represents the grain size distribution of particles across the disk for a snapshot taken at $\approx 1$ Myr from the beginning of our simulations. First, in the models where the radial drift is included (top left panel), without any kind of reduction or pressure bumps, no mm-sized grains are found at $> 10$ au from the star because of the radial drift of these particles. 
The effects of the radial drift mechanism are evident in the comparison with the model in which radial drift was neglected (top right panel). In this case, grains grow to sizes of $> 1$ mm all the way to about 100 au. At this time, at any given distance from the brown dwarf within 100 au, the size of the largest grains is defined by the \textit{fragmentation barrier} $a_{\rm{frag}}$ \citep{Birnstiel:2010}, as grains do not drift radially. In this regime, even larger grains would be obtained in disk regions with lower viscosity and higher fragmentation velocities than those considered in Fig.~\ref{fig:dust_evolution}, as $a_{\rm{frag}} \propto v_{\rm{frag}}^{2} \alpha_{\rm{v}}^{-1}$. 

Relative to the model with unimpeded drift, reducing the radial drift by 40 or 60\% (the latter case shown in the bottom left panel in Fig.~\ref{fig:dust_evolution}) does not show significant differences in the dust density distribution at $\approx 1$ Myr. In the models with strong sinusoidal bumps ($A = 0.6$, bottom right panel), particles with mm-cm sizes are instead efficiently trapped up to radii of about 100 au.


To summarize the main results of our calculations, disk models in which the radial drift mechanism has been either completely made inefficient, or significantly slowed down by the presence of strong gas pressure bumps are necessary to explain the presence of mm-sized grains in the outer regions of the 2M0444 disk.




\subsection{The excess emission at 1.36 cm}
\label{sec:excess}

\begin{figure*}[th!]
\centering
\includegraphics[scale=0.7]{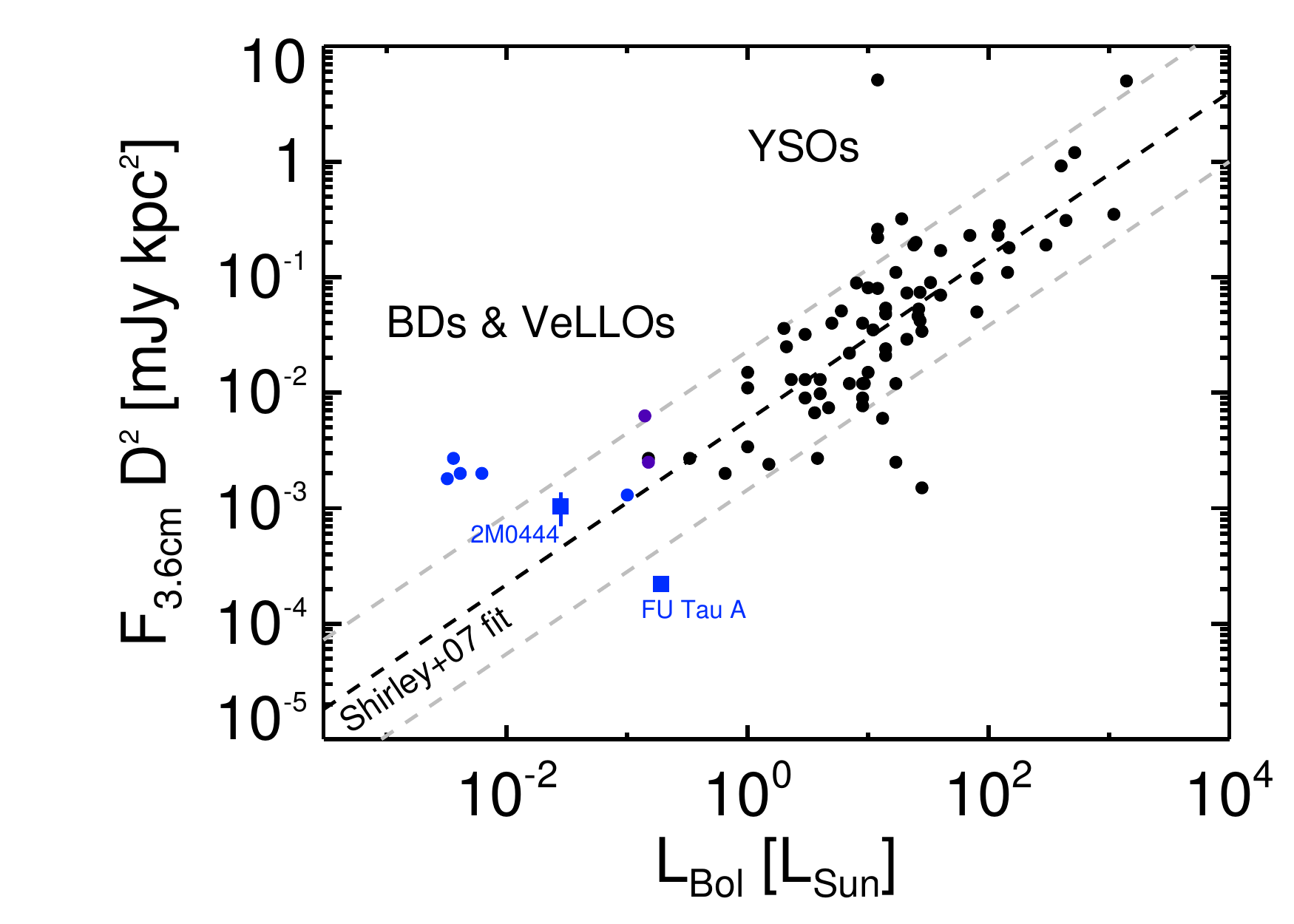}   
\caption{Radio luminosity at 3.6 cm vs bolometric luminosity for YSOs, young BDs and VeLLOs, after \citet{Morata:2015} and \citet{Rodriguez:2017}. Black dots are from the cm-wave surveys of YSOs by \citet{Anglada:1995}, \citet{Furuya:2003}, and \citet{Rodmann:2006}, with bolometric luminosities for the Rodmann et al. survey from \citet{Andrews:2013},  purple dots are for the VeLLOs from \citet{Andre:1999} and \citet{Shirley:2007}, blue dots are for the Class 0/I proto-brown dwarfs from \citet{Palau:2014}, \citet{Morata:2015} and \citet{Forbrich:2015}. Blue squares represent the two Class II brown dwarf disks, FU Tau A \citep{Rodriguez:2017} and 2M0444 (this work). For 2M0444, the vertical bar reflects the range of values derived from the extrapolation of the excess emission at 1.36 cm assuming a range of spectral indices between $-0.1$ and 0.6. The best-fit relation, and 1$\sigma$ uncertainty, derived for a sample of YSOs by \citet{Shirley:2007} are shown by dashed lines.}
\vspace{0.2cm}
              \label{fig:radio_bol}              
\end{figure*}

As described in Section~\ref{sec:results}, the measured flux density at 1.36 cm indicates the presence of ionized gas emission in excess of the dust thermal emission. This excess emission at cm wavelengths has been observed in several disks around young PMS stars \citep{Rodmann:2006,Ubach:2012,Pascucci:2014,Ubach:2017}.
For these disks, the spectral indices measured at cm wavelengths indicate free-free emission from a wind or jet, and chromospheric emission associated with stellar activity as the physical mechanisms responsible for this excess. In the case of young brown dwarfs, ionizing UV radiation can be produced by hot shocked gas accreting on the brown dwarf. \citet{Herczeg:2008} derived a temperature of $\approx 8200$ K for the shocked accreting gas by fitting the Balmer continuum observed from 2M0444.

Figure~\ref{fig:radio_bol} shows the relation between radio luminosity at 3.6 cm and bolometric luminosity for Young Stellar Objects (YSOs), Very Low Luminosity Objects (VeLLOs) and young BDs, spanning about 6 orders of magnitude in luminosity. The median of the spectral index measured between 3 and 6 cm for  protostars is $\approx 0.5$~\citep{Shirley:2007}, in line with partially optically thick free-free emission from a wind or jet with a $1/r^{2}$ density gradient \citep{Panagia:1975,Wright:1975,Reynolds:1986}.
This correlation likely reflects the fact that more luminous protostars drive more powerful ionizing winds and jets, and are likely surrounded by more massive circumstellar material, especially for the youngest Class 0/I YSOs which are still embedded in their parental envelope  \citep{Curiel:1987,Curiel:1989,Hsieh:2016}.

This relation has been recently extended to very low luminosities, $L_{\rm{bol}} \lesssim 0.1~L_{\odot}$, via cm-wave observations of very young Class 0/I proto-BDs \citep{Morata:2015,Forbrich:2015} and one Class II BD disk, FU Tau A \citep{Rodriguez:2017}, although a mass \textit{above} the hydrogen burning limit has been proposed for this object using models of magnetic stars to explain its observed high luminosity \citep{Stelzer:2013}.

In order to add 2M0444 to the plot in Fig.~\ref{fig:radio_bol}, we extrapolated to 3.6 cm the excess emission measured at 1.36 cm. We adopted a range for the cm-wave spectral index between $-0.1$ (valid for optically thin free-free emission) and $0.6$ (free-free emission from a wind or jet with a $1/r^2$ density structure), which includes the values measured for nearly all the more massive YSOs \citep{Shirley:2007}.

Fig.~\ref{fig:radio_bol} shows that the radio emission of the 2M0444 disk lies on the relation inferred by \citet{Shirley:2007} for YSOs. Simultaneous observations at longer wavelengths are necessary to confirm that the 1.36 cm excess emission is indeed due to free-free emission, rather than, e.g., synchrotron emission from electrons accelerated by the sub-stellar magnetic field. If so, the fact that 2M0444 lies so close to the relation derived for more massive young stars would be a further indication in favor of the scale down stellar-like scenario for the formation of this brown dwarf.
Furthermore, being the inferred radio luminosity for 2M0444 almost an order of magnitude higher than for the only other Class II BD disk detected at cm wavelengths, FU Tau A, our observations show that young brown dwarfs can be very active also at this early stage of their evolution.



\section{Conclusions}
\label{sec:conclusions}

We presented new VLA observations for the dust continuum emission at 6.8, 9.1 mm and 1.36 cm for the young brown dwarf 2M0444. The main results are as follows.

\begin{itemize}

\item Dust thermal emission from the 2M0444 disk was detected at 6.8 and 9.1 mm, representing the first detection of dust emission from a young BD disk at these long wavelengths.  The low value of the spectral index, $\alpha_{\rm{3.2 - 9.1mm}} \approx 2.37 \pm 0.32$, indicates that particles even larger than those invoked by the results of previous ALMA observations \citep{Ricci:2014} are present in the outer regions of the 2M0444 disk;

\item The disk sub-mm/mm SED shows a steepening at longer wavelengths. This feature of the SED can be explained solely by the steepening of the Planck function at long wavelengths for dust thermal emission at temperatures $T \approx 10 - 13$ K, in line with the disk model proposed by \citet{Ricci:2014}. However, the predictions of dust models suggest that some contribution is likely also due to a steepening of the spectral index $\beta$ of the dust opacity at wavelengths longer than $\sim 1$ mm;

\item Models of dust evolution show that the radial drift mechanism affecting solids in a gaseous environment has to be either completely made inefficient, or significantly slowed down by strong gas pressure bumps in order to explain the presence of mm/cm-sized grains in the outer regions of the 2M0444 disk. If this is done by radial gas pressure bumps, the amplitude of the bumps relative to the unperturbed gas structure has to be of at least $\sim 60\%$. The same models show also that the fragmentation barrier lies above sizes of $\approx 1$ mm in the outer disk regions;

\item The flux measured at 1.36 cm is dominated by ionized gas emission. The inferred radio luminosity for this emission is in line with the relation between radio and bolometric luminosity valid for YSOs \citep{Shirley:2007}, and extrapolated down to the very low luminosity of the 2M0444 brown dwarf. 

\end{itemize}

Future observations at high angular resolution ($< 0.1''$) at sub-mm and mm wavelengths with ALMA and the VLA will allow an in-depth investigation of the spatial distribution of the mm/cm-sized particles in the 2M0444 disk. In the case of the model with radial pressure bumps presented in Section~\ref{sec:dust} (Fig.~\ref{fig:dust_evolution}, bottom right panel), annular rings would have radial separations $\simgreat~10$ au, or $\simgreat~0.07''$ at the distance of 2M0444, in the disk outer regions.

Finally, follow-up multi-epoch observations at wavelengths longer than 1.36 cm will constrain the spectral index of the cm-wave emission as well as its time variability, and determine its physical nature. According to the models by \citet{Reiners:2010}, magnetic fields $\simgreat~1$ kG are expected for brown dwarfs with the mass and age of 2M0444. This level of magnetic field can be detected via modeling of the profile of spectral lines in high resolution spectra at NIR wavelengths~\citep[e.g.,][]{Reiners:2009}.

\appendix

Given the low luminosity of brown dwarfs, heating in the outer regions of their disks can be dominated by the diffuse radiation field from the interstellar medium (ISM), rather than by the radiation from the brown dwarf itself. 
In this Appendix we present a physical model that attempts to quantify the impact of this external heating in the case of the 2M0444 disk, for different values of the interstellar radiation flux.

For this calculation, we use the radiative transfer and ray tracing modules of the DALI code \citep{Bruderer:2012,Bruderer:2013}. 
For the surface density we consider two different parametrizations, both consistent with the CARMA and ALMA interferometric visibilities of the 2M0444 disk \citep{Ricci:2013,Testi:2014}. The first one is the power-law model described in Section~\ref{sec:dust}, i.e. a surface density radial profile $\Sigma_{\rm{dust}}(R) \propto R^{-p}$ with an exponent $p = 1.65$, truncated at an outer radius of 139 au, and with a total disk mass (gas$+$dust) of 1.3 $M_{\rm{Jup}}$ (gas-to-dust mass ratio of 100).

The second parametrization is the self-similar radial profile \citep{Pringle:1981}:

\begin{equation}
\Sigma_{\rm dust}(R)  = \Sigma_{\rm c} \left( \frac{R}{R_{\rm c}} \right)^{-\gamma} \exp{\left[ -\left( \frac{R}{R_{\rm c}} \right)^{2-\gamma} \right]}.
\end{equation}

For both parametrizations, at any given radius the total dust density is divided into two populations of dust particles with the same radial dependence, $\Sigma_{\rm dust}(R) = \Sigma_{\rm small} + \Sigma_{\rm large}$, but different grain sizes and vertical distributions to account for the vertical settling mechanism. Following~\citep{Testi:2014}, we adopt values of $\gamma = 1.4$, $R_{\rm{c}} = 50$ au.
For our reference model, the mass ratio between the two populations is set to $\Sigma_{\rm large}/\Sigma_{\rm small}=0.85$, i.e. at any location in the disk $85\%$ of the dust mass is in the \textit{large} population. The small population is made of particles with sizes ranging between $50\,\AA$ and $1\,\mu$m, whereas the large one of particles with sizes ranging between $1\,\mu$m and $1\,$mm. Both the populations assume a slope of $q = 3.5$ for the grain size distribution, $dn(a)/da \propto a^{-q}$, similar to the value constrained for the ISM~\citep{Mathis:1977}.

The density structure of the small grains assumes hydrostatic equilibrium and vertical isothermality, leading to:
\begin{equation}
\rho_{\rm small} (R,z)= \frac{\Sigma_{\rm small}(R)}{ \sqrt{2\pi} H} \exp { \left[ -\frac{1}{2} \left( \frac{z}{H} \right)^2 \right] },
\end{equation}
where $H$ is the scale-height of the disk. As for the large grains, we account for vertical settling reducing the scale-height of the second population by a factor $\chi=0.2$:

\begin{equation}
\rho_{\rm large} (R,z)= \frac{\Sigma_{\rm large}(R)}{ \sqrt{2\pi} \chi H} \exp { \left[ -\frac{1}{2} \left( \frac{z}{\chi H} \right)^2 \right] }.
\end{equation}
The scale-height of the disk scales as $H\propto R^{1.25}$, with $H/R=0.21$ at $R_{\rm c}$, as derived by \citet{Ricci:2014} under the assumption of vertical hydrostatic equilibrium and after subtracting the spatially constant $T_{\rm{ex}}$ term \citep[see Section 4 in][]{Ricci:2014}. The 2D spatial grid consists of 150 grid points in the radial direction, logarithmically sampled between $0.01\,$AU and $150\,$AU, and of 80 grid points in the vertical direction, sampled linearly between the disk mid-plane and 8 scale-heights above it.

The dust opacities of the two populations are computed using a standard ISM dust composition following \citet{Weingartner:2001}. The mass extinction coefficients are calculated using Mie theory with the \verb!miex! code \citep{Wolf:2004} and optical constants by \citet{Draine:2003} for graphite and \citet{Weingartner:2001} for silicates. The input sub-stellar spectrum is set by the observed stellar properties, in particular effective temperature, luminosity (see Section~\ref{sec:intro}) and mass accretion rate~\citep[$\dot{M}_{\rm{acc}} \approx 9 \times 10^{-12}~M_{\odot}$ yr$^{-1}$,][]{Herczeg:2008}, as constrained for the 2M0444 brown dwarf. 

The external UV field is presented in units of $G_0$, where $G_0 \sim 2.7\times10^{-3}$\,erg\,s$^{-1}$\,cm$^{-2}$ is the average UV interstellar radiation field between $911\,\AA$ and $2067\,\AA$ \citep{Draine:1978}. In star forming regions, the environmental UV field can range between a few $G_0$ in low mass regions, up to $>10^5\,G_0$ in the proximity of O stars in massive clusters as in the Orion Nebula Cluster \citep[e.g.][]{Fatuzzo:2008}. As an example for low mass forming regions, \citet{Cleeves:2016} estimated an external field of $\sim 4\,G_0$ from thermo-chemical models of gas observations in IM Lup, then corroborated by hydro-dynamical models of external photoevaporation \citep{Haworth:2017}. 

The number of photon packages used in the radiative transfer is $3\times10^7$ both for the photons from the central star and for the photons from the environment, which are emitted from a virtual sphere.

\begin{figure*}[h!]
\centering
\includegraphics[scale=0.6]{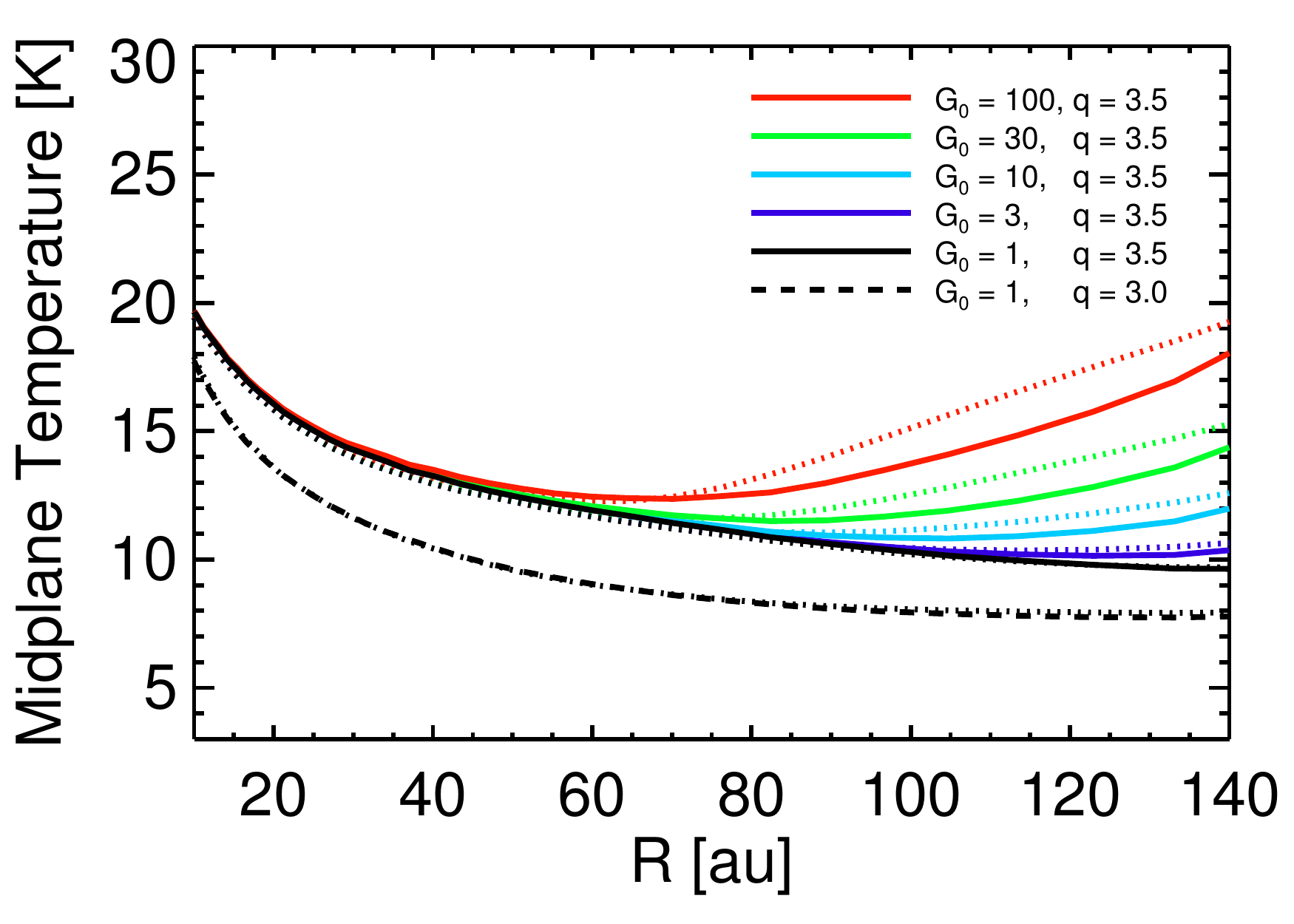}   
\caption{Radial profiles of the dust temperature in the disk midplane as derived by our radiative transfer calculations. The fluxes considered for the external UV interstellar field are shown with different colors as labelled. Solid and dotted lines refer to the calculations for the power-law and self-similar radial profiles for the dust density, respectively, and for our reference model for the grain size distribution. Dashed line represents the model with a grain size distribution with a slope $q = 3.0$ (see text).}
              \label{fig:r_temp}
\end{figure*}

Figure~\ref{fig:r_temp} shows the radial profiles of the dust temperature in the disk midplane for different values of the flux of the external UV field (lines with different colors). Deviations in the temperature profiles for the power-law (solid lines) and self-similar (dotted) models are confined to $\simless~1-2$ K at the same radius.
The temperature profiles obtained for low-to-moderate values of $1 \leq G_0 \leq 3$ resemble very closely the temperature profile from the \citet{Ricci:2014} model for 2M0444. Only for stronger UV fields, with $G_0 \simgreat 10$, the dust temperature shows an inverted trend: the temperature rises further from the central object because of the dominant heating from the external radiation field.

In this plot we also test the variation of temperature with the dust opacity. We do this by considering a disk with the same structure as described above, but with different dust opacities obtained by a rather extreme choice of a top-heavy grain size distribution, with a slope $q = 3.0$ and a ratio $\Sigma_{\rm large}/\Sigma_{\rm small}=0.99$. Compared with the reference model, the temperature is lower by just a couple of degrees, and remains always above 8 K even with a very low flux of $G_0 = 1$ for the external UV field.

\acknowledgments

The National Radio Astronomy Observatory is a facility of the National Science Foundation operated under cooperative agreement by Associated Universities, Inc.
T.B. acknowledges funding from the European Research Council (ERC) under the European Union's Horizon 2020 research and innovation programme under grant agreement No 714769.

\end{document}